\documentclass[12pt]{article}
\usepackage{epsf}
\begin{document}
\def\be{\begin{equation}}
\def\ee{\end{equation}}
\def\ba{\begin{array}{l}}
\def\ea{\end{array}}
\def\bea{\begin{eqnarray}}
\def\eea{\end{eqnarray}}
\def\eq#1{(\ref{#1})}
\def\fig#1{Fig \ref{#1}} 
\def\wgnc{\bar{\wedge}}
\def\del{\partial}
\def\der{\overline \del}
\def\wg{\wedge}
\def\bull{$\bullet$}
\def\gap{\vspace{10ex}}
\def\tgap{\vspace{3ex}}
\def\sgap{\vspace{5ex}}
\def\lgap{\vspace{20ex}}
\def\half{\frac{1}{2}}
\def\pto{\vfill\eject}
\def\gst{g_{\rm st}}
\def\tC{{\widetilde C}}
\def\z{{\bar z}}
\def\o{{\cal O}}
\def\J{{\cal J}}
\def\S{{\cal S}}
\def\X{{\cal X}}
\def\N{{\cal N}}
\def\A{{\cal A}}
\def\H{{\cal H}}
\def\D{{\tilde D}}
\def\d{{\cal D}}
\def\re#1{{\bf #1}}
\def\nn{\nonumber}
\def\nl{\hfill\break}
\def\ni{\noindent}
\def\bibi{\bibitem}
\def\c#1{{\hat{#1}}}
\def\eps{{\epsilon}}
\pretolerance=1000000
\begin{flushright}
CERN-TH/2000-130\\
TIFR/TH/00-33\\
May 2000\\
\end{flushright}
\begin{center}
\vspace{2 ex}
{\large\bf Gauge Theory on a Quantum Phase Space }
\\
\vspace{3 ex}
Luis Alvarez-Gaume $^*$ and Spenta R. Wadia $^*$ $^\dagger$ \\
~\\
{\sl Theory Division, CERN, $^{*}$} \\
{\sl Ch-1211, Geneva 23, SWITZERLAND.}\\
\vspace{1 ex}
{\sl and}\\
\vspace{1 ex}
{\sl Department of Theoretical Physics,$^{\dagger}$}\\
{\sl Tata Institute of Fundamental Research, }\\
{\sl Homi Bhabha Road, Mumbai 400 005, INDIA.} \\

\vspace{10 ex}
\pretolerance=1000000
\bf ABSTRACT\\
\end{center}
\vspace{1 ex}
In this note we present a operator formulation of gauge theories in a
quantum phase space which is specified by a operator algebra. For
simplicity we work with the Heisenberg algebra. We introduce the
notion of the derivative (transport) and Wilson line (parallel
transport) which enables us to construct a gauge theory in a simple
way. We illustrate the formulation by a discussion of the Higgs
mechanism and comment on the large N masterfield.

\vfill

\clearpage

\vspace{8ex}

\section{\large Introduction} 

Recently there has been much interest in the study of non-commutative
(nc) geometry in the context of quantum field theory (including gauge
theories) and string theory $\cite{doug,dougconnes,doughull,Ardalan,Krogh,garcia,chu,schom,froh}$ There
are several directions of exploration. To begin with let us cite some
examples of nc geometry.

\begin{enumerate}

\item Witten's open string field theory is formulated with a
associative, non-commutative product $\cite{witten1}$. 

\item The BFSS matrix model formulation of M-theory. Here space
coordinates are matrices and hence there is a natural associative,
non-commutative product. If we express the Hamiltonian of this model
in terms of variables that describe 2-branes the associative,
non-commutative matrix algebra is reflected in the 2-brane world
volume being a 2-dim. phase space. $\cite{bfss,nicolai}$.

\item World volume theories of brane systems in the presence of
certain moduli fields like the Neveu-Schwarz $B_{NS}$ (in the
Seiberg-Witten limit) become field theories where ordinary
multiplication of fields is replaced by the Moyal star product in a
phase space. $\cite{witten2}$ .

\item In the study of the large $k$ limit of the $SU(2)$ WZW model
describing open strings moving on a group manifold one encounters the
coadjoint orbits of $SU(2)$. These are 2-spheres with a natural
symplectic form (called fuzzy 2-spheres.) The radius of the coadjoint
orbit is $j \leq k/2$.  $\cite{schom1}$

\item The fuzzy 2-sphere also occurs in the Polchinski-Strassler
description of bulk geometry in the presence of relevant perturbations
of the $D_3$ brane system $\cite{polch, sandip}$.

\item The c=1 matrix model is exactly formulated as a particular
representation of the $W_{\infty}$ algebra, that comes from
1-dim. non-relativistic fermions. The coadjoint representation is
carried by the phase space density $u$ of fermions constrained by the
equation $u\star u=u$, which is a quantum statement of fermi
statistics $\cite{SRW}$. Collective field theory $\cite{jevicki}$ describes the classical geometry
limit where the constraint involves the ordinary product $u^{2}=u$.

\item The gauge invariant description of 2-dim. $U(N)$ QCD with
fermions also leads to a specific representation of the $W_{\infty}$
algebra. Here too the coadjoint representation is described by the
equation $M*M=M$ where M represents the gauge invariant Wilson line
between two quarks. The Regge trajectory of mesons appear as solutions
of the small fluctuation equation around the large N classical
solution of $M*M=M$ $\cite{SRW2}$.

\end{enumerate}

The examples cited above share one important feature
in common: they are theories whose fields are valued in a ${\it
quantum~ phase~ space}$ rather than on a manifold. It is presently not clear how Witten's string field theory fits into this framework. \footnote {Witten has recently made progress in this direction. See the note added
at the end of this paper.}

The quantum phase
space is specified by a operator algebra. By virtue of this, these
fields become operators and act in an appropriate Hilbert space. If
one considers the coadjoint representation of the operator algebra
then one can have a correspondence between the operator valued fields
and classical functions on the coadjoint orbit. For the simplest
example of the Heisenberg algebra the coadjoint orbit corresponds to
the familiar phase space. The composition law for functions on the
phase space is the Moyal star product. This point is briefly reviewed
in the next section and is called the Weyl-Moyal correspondence.

Besides the fact that space-time is non-commutative at the fundamental
level, as suggested by the Matrix model formulation of M-theory, 
(see also $\cite{yoneya}$)
this new type of field theory has many interesting properties. One very
significant property is the IR/UV connection $\cite{seib1}$.  Such a
connection is indeed novel and seems to signal a breakdown of
decoupling. As has been suggested it may be useful in understanding
the cosmological constant problem $\cite{seib1}$. Another important
property of a nc theory is the role it plays in the resolution of
singularities $\cite{nekra,seib2,srw3}$. Another use can be in
understanding the large N limit of gauge theories.

With these and other applications in mind it is important to
understand these theories in various ways. In this note we formulate a
gauge theory on a quantum phase space. Our main point is to give a
formulation directly in the language of operators, using simple rules
that physicists are familiar with. For simplicity we discuss the
Heisenberg algebra and show that the operator formulation of the gauge
theory corresponds to a formulation on the coadjoint orbit (phase
space) in terms of the Moyal star product.

In section 2. we introduce the basic notion of a derivative operator
which can translate operators. We also introduce the notion of
operator valued forms, the exterior derivative, the wedge product and the Dirac operator. In section 3. we discuss the
Weyl-Moyal correspondence. In section 4. we introduce the notion of
parallel transport of operators using the Wilson line. We also
introduce the corresponding gauge field , gauge transformations, 
field strength, action and instanton number. In section 5. we
define the Wilson loop. In section 6. comment on the quantum theory.
In section 7. we discuss the Higgs mechanism and 
in section 8. we make some remarks on the large N master field.

\section{\large The Derivative Operator on a Quantum Phase Space}

A quantum phase space
is specified by an operator algebra. This algebra can have finite or
infinite number of generators. Examples of finite number of generators
are the Heisenberg algebra and Lie algebras of compact and non-compact
groups. The algebras with infinite number of generators are eg. the
Kac-Moody algebra and the Virasoro algebra. One can and should include
super-algebras to this list.

To simplify matters we restrict ourselves to the Heisenberg algebra.
\be
\label{1.1}
[X_i, X_j]=iI\theta_{i,j}
\ee
where , $I$ is the identity operator, $i,j = 1,2,..,2d$, and
$\theta_{i,j}$ is a real anti-symmetric, invertible matrix, with
inverse $\theta^{-1}_{i,j}$.  
We have to specify the Hilbert space on which these operators act. For
our present purposes we can take this to be the space of delta-function 
normalizable functions in d-dimensions.

Using the basic operators \eq{1.1}
we can construct other operators in terms of polynomials of $X_{i}$
over the complex numbers. Instead of a polynomial basis we can more 
fruitfully use the Weyl basis defined by the exponential operators
\be
\label{1.2}
g(\alpha)=\exp{i\alpha_iX_i}
\ee
In this way we can introduce complex, self-adjoint and unitary
operators in the Hilbert space.

Since we would like to develop an operator calculus, the first thing
that we should do is to define the derivative operator and the notion
of translations in the space of operators. We define the derivative by,
\be
\label{1.3}
\der_i=\theta^{-1}_{i,j}adX_j
\ee
where the adjoint action is defined by $(adA)B=[A,B]$.
This derivative operator has a number of important properties which we list:

\ni 1. $\der_i$ is anti-hermitian and linear,
\bea
\label{1.4a}
\der_i^{\dagger} & = & -\der_i \nn \\
\der_i(a_1\o_1 + a_2\o_2) & = & a_1\der_i\o_1 + a_2\der_i\o_2 \nn \\
\eea 

\ni 2. $\der_i$ satisfies the Leibniz rule,
\be
\label{1.4b}
\der_i(\o_1\o_2)=(\der_i\o_1)\o_2 + \o_1(\der_i\o_2)
\ee

\ni 3. $\der_i$ commute amongst themselves,
\be
\label{1.4c}
[\der_i, \der_j]=0
\ee

\ni 4. The commutative property of the derivative enables us to introduce the notion of an exterior derivative that acts on
operator valued forms.
The operator valued n-form and its exterior derivative are defined by,
\bea
\label{1.4d}
O^{(n)} &=& \sum \o_{i_1,..,i_n}dy_{i_1}\wg...\wg dy_{i_n}
\nn \\
dO^{(n)} &=& \sum \der_{i_1}\o_{i_2,..,i_n}dy_{i_1}\wg...\wg dy_{i_{n+1}} \nn \\
\eea
In the above $dy_i$ are real 1-forms. The commutative property of the
derivative clearly implies that $d^{2}=0$. 

Using the above definition of the operator
valued n-form we can introduce the notion of a non-commutative 
wedge product 
\be
O^{(n)}\wgnc O^{(m)} = C_{n,m} \sum \o_{i_1,.,i_n}\o_{i_{n+1}},.,i_{n+m}
dy_{i_1}\wg...\wg dy_{i_{n+m}} \\
\ee
$C_{n,m}$ is a normalization constant. The above definition of the
wedge product is associative. One can also define $ ^{\star}O^{(n)}$  
the Poincare dual of
$O^{(n)}$ in the standard fashion using the totally antisymmetric $\eps$
tensor in $2d$ dimensions.

\ni 5. $\der_i$ is the generator of translations 
in the following sense,
\be
\label{1.4e}
\exp{ ia_i\der_i}\,\o(X)=\o(X+Ia)
\ee

\ni 6. The integral of an operator is defined by its trace in the
corresponding Hilbert space. Then using the trace formula
$TrA[B,C]=Tr[A,B]C$ we have the formula for integration by parts,
\be
\label{1.4f}
Tr\o_1(\der_i\o_2) = - Tr(\der_i\o_1)\o_2
\ee
We note that there is no `surface term'.

\ni 7. We can introduce the Dirac operator $\not\der=\gamma_i\der_i$, where
$\gamma_i$ are the standard Dirac gamma matrices. Note that 
$(\not\der)^2=\der_i\der_i$.

\par

\underline{The Landau Condition}

The defining equation for the derivative operator can be understood in 
a more physical setting by introducing additional momentum operators $P_i$
and extending the algebra,
\be
[P_i, P_j] = 0  
\ee
\be
[P_i, X_j] = {-i\delta_{ij}}I
\ee
and introducing the constraint,
\be
adP_i = \der_i  = \theta^{-1}_{i,j}adX_j
\ee
On states in the Hilbert space this equation becomes the Landau 
constraint $\cite{kogan,Iso,sakita,lenny}$,
\be
(P_i - \theta^{-1}_{i,j}X_j)|\Psi >=0
\ee
which implies for such states the uncertainty principle,
\be
\delta X_i \theta^{-1}_{i,j} \delta X_j \geq 1
\ee

\section{\large Weyl-Moyal Correspondence}
The Weyl-Moyal (WM) correspondence is best understood in terms  of the operators
$g(\alpha)=\exp{i\alpha_iX_i}$. 
Using \eq{1.1} these operators satisfy the Heisenberg-Weyl algebra,
\be
\label{2.1}
g(\alpha)g(\beta)=\exp{(\frac{1}{2}\theta_{i,j}\alpha_i\beta_j)} 
g(\alpha + \beta)
\ee

The WM correspondence is given by
\be
\label{2.2}
\o(X)=\int d^{2d}\,\alpha \,g(\alpha, X)\widetilde {O}(\alpha)
\ee
$\widetilde O$ is the Fourier transform of $O$.
Using (\ref{2.1}) we can derive correspondence between the operator product 
and the star product,
\bea
\label{2.3}
\o_1\o_2 &=& \int{ d^{2d}\,\alpha \,g(\alpha, X)\widetilde {O}
_{12}(\alpha)} \nn \\
O_{12}&=& O_1\star O_2 \nn \\
O_1\star O_2 &=& \exp(i\frac{1}{2}\theta_{i,j}\frac{\del}{\del\xi _i}\frac{\del}{\del\eta _j})
O_1(x+\xi)O_2(x+\eta)\mid _{\xi=\eta=0}
\eea

Using the WM correspondence, we can easily prove the correspondence,
\bea
\label{2.4}
\der_i\o(X) & \Longrightarrow & \del_iO(x) \nn \\
\o(X)_1\o(X+Ia)_2 & \Longrightarrow & O(x)_1\star O(x+a)_2 \nn \\
\eea

\section{\large Parallel Transport of Operators, Wilson Lines and Gauge Fields}

In section 2. we introduced the notion of translating
operators. In this section we introduce the notion of parallel transport
and connection. For convenience of presentation we will suppress the $U(N)$ indices carried by the various operators.

Consider the set of operators ${\Phi} (X)$, which transform under the
right action of the group of unitary operators $\{ \Omega(X)\}$ 
\be
{\Phi} (X) \rightarrow {\Phi} (X){\Omega}(X)
\ee
Then clearly 
\be
{\Phi} (X){\Phi} (X+Ia)^{\dagger}
\ee
is not gauge invariant under the gauge transformations
\bea
{\Phi} (X) & \rightarrow & {\Phi} (X){\Omega}(X) \\
{\Phi} (X+Ia) & \rightarrow & {\Phi} (X+Ia){\Omega}(X+Ia)
\eea

The standard way to form a gauge invariant operator is to introduce 
the Wilson line $U(X,X+Ia)$, with gauge transformation
\be
U(X,X+Ia) \rightarrow {\Omega}(X)^{\dagger}{U}(X,X+Ia)
{\Omega}(X+Ia)
\ee
and the property that $U(X,X)=I$ and $U(X,X+Ia)^{\dagger}=U(X+Ia,X)$.
The operator 
\be
{\Phi} (X)U(X,X+Ia){\Phi} (X+Ia)^{\dagger}
\ee
is gauge invariant.

Now a similar construction is possible for those operators $\Psi (X)$
which transform under the left action of the group of unitary operators,
\be
{\Psi} (X) \rightarrow {\Omega}(X){\Psi} (X)
\ee
In this case the gauge invariant operators are given by,
\be
{\Psi} (X){U}(X,X+Ia)^{\dagger}{\Psi} (X+Ia)^{\dagger}
\ee

\par

\underline{The Connection}:

Using the Wilson line for infinitesimal $a_i=\eps_i$ we can
introduce the definition of the operator valued connection,
\be
{U}(X,X+I\eps)=\exp (i\A_i(X)\eps_i)
\ee
The operator $\A_i(X)$ is Hermitian and the gauge transformation of the Wilson line
implies the gauge transformation of the operator valued gauge field.
\be
\A_i(X)  \rightarrow {\Omega}(X)^{\dagger}(\A_i(X) - i\der_i){\Omega}(X) 
\ee

\underline{Covariant Derivative and Field Strength}:

The operator covariant derivative and the field strength are defined by,
\bea
\d_i &= & -i\der_i + \A_i(X) \\
{\cal F}_{ij}(X) & = &  i[\d_{i}, \d_{j}]
\eea
Both the covariant derivative and the field strength are gauge
covariant and the Jacobi identity for the covariant derivative
implies the Bianchi identity for the field strength,
\be
[\d_i,\,{\cal F}_{jk}]+[\d_j,\,{\cal F}_{ki}]+[\d_k,\,{\cal F}_{ij}]=0
\ee

\par

\underline{The Action and Equations of Motion}:

The gauge invariant action is given by
\bea
S & = & \frac{1}{4g^2}Tr({\cal F}(X)\wgnc ^{\star}{\cal F}) \nn \\
& = &\frac{1}{4g^2}Tr({\cal F}_{ij}(X){\cal F}_{ij}(X))
\eea
In the above we have chosen the euclidean metric. From the above action
we can easily derive the equations of motion by requiring the action to
be stationary w.r.t the variation $\A_i(X)\rightarrow \A_i(X)+ \delta
\A_i(X)$,
\be
\d_i\,{\cal F}_{i,j}(X)=0 
\ee

We can also define the nc version of the instanton number
\bea
\label{inst}
I & = & Tr({\cal F}(X)\wgnc {\cal F}) \nn \\
& = & Tr({\cal F}_{i,j}(X){\cal F}_{k,l}(X)\eps_{ijkl})
\eea
(In the above formulas the trace can also includes a trace over $U(N)$.)

It is also possible to introduce the following operator current,
\be
\J_i = \epsilon _{ijkl}(\A_j\der_k\A_l + i\frac {2}{3}\A_j\A_k\A_l)
\ee
so that (\ref{inst}) can be written as,
\be
I=4\,Tr(\der_i\J_i)
\ee

Using the WM correspondence it is easy to prove that the operator
formulation given above goes over into a gauge theory formulated in terms
of a real connection, and the star product, e.g.
\bea
\A_i(X) & \longrightarrow & A_i(x) \nn \\
{\cal F}_{ij}(X) & \longrightarrow & \del_iA_j -\del_jA_i +
i(A_i\star A_j- A_j\star A_i) \nn \\
\eea

In the operator formulation the gauge group is generated by all
unitary operators.  The generators of this gauge group are the set of
all Hermitian operators: $\{H(X)\}$, and $\{\Omega(X)=\exp i{\cal
H}(X)\}$.  Expressing the exponential as a series we can easily obtain
the correspondence for the gauge group,
\be
\exp i{\cal H}(X) \longrightarrow (1+H(x)+\frac{1}{2}H(x)\star H(x) +
\frac{1}{6}H(x)\star H(x)\star H(x)...)
\ee

\section{\large The Wilson Loop }
Let us now present the expression for the Wilson loop operator.
Consider a curve $\Gamma $ in $R^{2d}$ and divide it into $n\rightarrow \infty$ 
infinitesimal segments each denoted by a tangent vector $\eps^{m}_i$.
$i=0,1,2..n$ and  $\eps^{0}_i=0$. Since the loop is closed
we have $\sum{i=1}^{n}\eps^{m}_i=0$. The Wilson loop is composed of a product of 
Wilson lines around the curve,
\bea
W(\Gamma) & = & Tr\, \prod_{m=0}^{n}\, U(X+I\eps^{m},\,X+I\eps^{m+1}) \nn \\
&=& Tr\, \prod_{m=0}^{n}{\exp{i(\,\A_i(X+I\eps^{m+1})(\eps^{m}_i-\eps^{m+1}_i))}}
\eea
$W(\Gamma)$ is gauge invariant. It would be useful to see the
connection of this formulation with the reduced model of lattice gauge
theory $\cite{kitazawa,szabo}$.

\section{\large The Quantum Theory}

Until now we have been dealing with a classical theory whose fields
are defined on a `quantum phase space' specified by the matrix
$\theta_{i,j}$.  Quantization of this theory consists of studying
fluctuations whose strength is controlled by the gauge coupling. One
quantization procedure appeals to the WM correspondence and gives a
path integral prescription in which one integrates over the histories
of the gauge field $A_i(x)$.  Another approach to
quantization is to quantize the matrix elements of $\A_i(X)$ in a
coherent state basis $\cite{perelom}$.

\section{\large The Higgs Mechanism}

We now discuss `matter fields' in the nc gauge theory. For simplicity we discuss
matter fields $\Psi (X)$ with gauge transformation (only left action)
${\Psi} (X) \rightarrow {\Omega}(X){\Psi} (X)$.
The gauge invariant action is given by
\be
S= \frac{1}{4\,g^2}Tr({\cal F}_{i,j}(X){\cal F}_{i,j}(X)) + 
\frac{1}{2}Tr\,(\d_i\Psi)^{\dagger}(\d_i\Psi) + \\
\frac{1}{4}Tr({\Psi}^{\dagger}\Psi-a^2\,I)^2
\ee

To discuss the Higgs mechanism, we write $\Psi=U\H$, where $U$ is unitary and $H$
is Hermitian, and perform the gauge transformation to the unitary gauge 
$\Psi \rightarrow U^{\dagger}\Psi$.
In this gauge the potential term becomes $V=\frac{1}{4}Tr(\H^2-a^2\,I)^2$
and the ground state is a solution of the operator equation,
\be
\label{higgs}
\H^3=a^2\H
\ee
or equivalently
\be
H\star H\star H = a^2\,H
\ee
If we assume that $\H^{-1}$ exists then (\ref{higgs}) reduces
to
\be
\H^2=a^2\,I
\ee
Such operator equations were originally discussed in $\cite{SRW,SRW2}$
They have also been recently discussed in the context of soliton
solutions of nc field theories $\cite{gopa}$. See also
$\cite{mukhi,harvey}$ for subsequent applications. The minimum energy
solution is given by $\H = |a|I$, where we have absorbed a possible sign
ambiguity by a gauge transformation. This leads to a `mass term' for
the gauge field $|a|^2\,Tr(\A_i\A_i)$. It would be interesting to look
for vortex like solutions in these models.

\section{\large The large N Master Field}
Let us now consider the nc gauge theory with additional color gauge
group $U(N)$. In ref. $\cite{seib1, rajesh}$ the perturbation
expansion in the maximally non-commutative regime ($\theta \rightarrow
\infty$ ) was discussed. In this limit the leading term in the
perturbation expansion consists of planar graphs which have no theta
or N dependence except for an overall phase involving $\theta$ and a
multiplicative factor of $N^2$. Hence the problem of summing planar
diagrams of the nc gauge theory is mapped onto the problem of solving
the nc gauge theory in the $\theta \rightarrow \infty$ limit.

Let us choose $ \theta_{ij} = \theta \left( \begin{array}{cc} 0 & -1_d
\\ 1_d & 0 \end{array} \right) $ and write the gauge theory in terms
of the scaled operators $X_i \over \sqrt{\theta}$ or after the WM
correspondence, in terms of the scaled co-ordinates $x_i \over
\sqrt{\theta}$. If we require that the gauge field has the
transformation $A_i({x_i \over \sqrt{\theta}})= \sqrt{\theta}A_i(x)$,
then the action becomes 
\be 
S = {\theta^{2d-4 \over 2 } \over {4g^2}}
\int d^{2d}x tr( F_{ij} \star F_{ij}) 
\ee 
and the path integral is
given by \be Z = \int {\cal D}A_i(x) e^{-S} \ee For $2d>4$ and the
path integral is evaluated in the $\theta \rightarrow \infty$ limit by
the saddle point $ {\delta S \over \delta A_i(x)} = \nabla_i F_{ij} =
0$. Reverting back to the operator formalism, for $2d>4$, is given by
four $\theta \rightarrow \infty$ operators $ {\cal A}_i(x)$ which
satisfy the n.c YM equations 
\be 
{\cal D}_i {\cal F}_{ij} = 0 
\ee 
It
is reasonable that (38) are equations for the $U(\infty)$ master field
for $2d>4$. 

The situation in the most interesting dimension is certainly more
complicated and the Schwinger-Dyson equations approach $\cite{sch}$ may help. Finally
it would be interesting to make a connection of the large N limit 
of nc gauge theory with non-commutative probability theory as applied to the problem of the large N limit by Gopakumar and Gross $\cite{gopagross}$.

\tgap

{\bf Note added:}

After this work was completed we received $\cite{ambjorn}$ where an
approach similar to ours has been discussed.  However the derivative
operator discussed in this paper (equation 2.7 ) does not necessarily
commute in different directions.  Gross and Nekrasov
$\cite{grossnekra}$ have also presented the basic ingredients of the
nc gauge theory in the operator formulation. Their derivative operator
is identical to ours. Recently, Witten $\cite{witten-moyal}$ has shown
the emergence of the Moyal product in the large B-field limit of
string field theory.

It turns out that the geometrical construction of the non-commutative gauge theory presented here can be shown to be equivalent to the perturbation of the 
IKKT model \cite{IKKT} around the brane configuration defined by \eq{1.1}.
This point has been made in \cite{kitazawa, szabo}. We would like to 
thank R. Szabo and A. Dhar for pointing this out to us.

\newpage

{\bf Acknowledgement:}

One of us (SRW) would like to thank the CERN theory division for a
visit during which this work was done.  We would like to thank
Farhad Ardalan, Hessam Arfei, Ali Chamseddine, Avinash Dhar, Gautam
Mandal, Dileep Jatkar, Sandip Trivedi and K.P. Yogendran for useful
discussions.

\end{document}